%% file: conference_101719.tex
\documentclass[conference]{IEEEtran}
\IEEEoverridecommandlockouts
\usepackage{cite}
\usepackage{amsmath,amssymb,amsfonts}
\usepackage{algorithmic}
\usepackage{graphicx}
\usepackage{textcomp}
\usepackage{xcolor}

\usepackage{gensymb}
\usepackage{siunitx}
\usepackage{booktabs} 
\usepackage{caption}
\usepackage{subcaption}
\usepackage{subcaption}
\usepackage{float}
\usepackage{gensymb}
\usepackage{enumitem}
\usepackage{ragged2e}
\usepackage{mdframed}
\usepackage{tikz}
\usepackage{booktabs}
\usepackage{eurosym}
\usepackage{multirow}
\usepackage{makecell}
\usepackage{array}
\usepackage{url}
\usepackage{lineno}

\def\BibTeX{{\rm B\kern-.05em{\sc i\kern-.025em b}\kern-.08em
    T\kern-.1667em\lower.7ex\hbox{E}\kern-.125emX}}
\begin{document}

\title{Optimisation of Electrolyser Operation: \\ Integrating External Heat}

\author{\IEEEauthorblockN{Matthias Derez}
\IEEEauthorblockA{\textit{Department of Mechanical Engineering} \\
\textit{KU Leuven}\\
Leuven, Belgium \\
}
\and
\IEEEauthorblockN{ Alexander Hoogsteyn}
\IEEEauthorblockA{\textit{Department of Mechanical Engineering} \\
\textit{KU Leuven}\\
Leuven, Belgium \\
alexander.hoogsteyn@kuleuven.be}
\and
\IEEEauthorblockN{Erik Delarue}
\IEEEauthorblockA{\textit{Department of Mechanical Engineering} \\
\textit{KU Leuven}\\
Leuven, Belgium \\
erik.delarue@kuleuven.be}
}

\maketitle

\begin{abstract}
Integrating external heat into electrolysers can reduce the electrical power demand for carbon-neutral hydrogen production. Efficient operation requires detailed models that incorporate heat availability and its effect on startup costs. This paper advances existing operational models by endogenously modelling startup costs and direct heat integration, based on a piecewise linear approximation of the electrochemical equations. We analyse the impact of low- and high-temperature heat integration on the efficiency and profitability of hydrogen production for solid oxide and proton exchange membrane (PEM) electrolysis technologies. Our case study demonstrates that heat integration can significantly enhance the profitability of solid oxide electrolysis, increasing average daily profits by up to 23.3\%. For PEM electrolysis, the impact is more modest, with profit gains of 1.9–2.7\%. Despite this, the break-even hydrogen price for solid oxide electrolysis remains high at 5.5 EUR/kg, given an investment cost of 2300 EUR/kW. PEM electrolysis achieves breakeven at more competitive hydrogen prices of 3.5–4.5 EUR/kg, depending on lifetime assumptions and an investment cost of 900 EUR/kW.
\end{abstract}


\input{math}

%% main text
\section{Introduction}
\label{sec:indtroduction}
To reduce the amount of renewable electricity required for green hydrogen production, heat can be integrated into the electrolysis process. Depending on the technology heat can be added either directly to provide a share of the required enthalpy change to carry out the electrolysis reaction or indirectly in other processes in the balance of plant. Both approaches reduce the required amount of electrical energy and improve the profitability. Efficiency improvements can be achieved using a variety of low carbon energy sources: geothermal energy \cite{geothermal}, biomass \cite{rivera} or nuclear energy \cite{zhang_75efficiency_2012, obrien_74high-temperature_2010}, among others.

Secondly, electrolysers can increase profitability by exploiting their flexibility by adapting their hydrogen production to fluctuating electricity prices and participate in ancillary services markets \cite{irena_128green_2020}. However, the electrical efficiency of an electrolyser depends on the operating temperature and current density. These operating parameters can change throughout the production period, introducing variations in electrical efficiency that are often not considered \cite{park_techno-economic_2024}. Advanced operational models, capable of capturing flexibility and variable efficiency, are necessary to accurately estimate the profits of an electrolyser \cite{baumhof_92optimization_2023}.

To accurately represent dynamic electrolyser operation, the effect of both the temperature and current density on efficiency must be considered. This can be done by incorporating the governing electrochemical Eqs., where the chemical processes that occur and the accompanying losses are described \cite{abdin_42modelling_2015, firtina-ertis_95thermodynamic_2022, zhao_84system_2021}. Alternatively, empirical electrochemical models can be used that approximate the relation \cite{ulleberg_67modeling_2003}. Such empirical models are available only for alkaline electrolysers but are nonetheless used to approximate the behaviour of proton exchange membrane (PEM) and solid oxide electrolyser (SOE) \cite{baumhof_92optimization_2023, varela_122modeling_2021, zheng_optimal_2022}. In this paper, operational models are developed which incorporate the original electrochemical Eqs. that relate the power consumed by the electrolyser to the operating temperature and current. To the author's knowledge, a piecewise linearisation of the original electrochemical model Eqs. has not been carried out in the literature. \cite{zheng_optimal_2022} have approximated the empirical model by Ulleberg using a single tangent plane to the power curve at nominal power. \cite{baumhof_92optimization_2023} have carried out a two-dimensional piecewise linearisation of \cite{ulleberg_67modeling_2003}, where the temperature dependency is neglected, limiting the accuracy of the model to discuss the effect of heat addition.
%The optimal operation of electrolysers is formulated as a mixed-integer bilinear programming model. Separate formulations were developed for the proton exchange membrane and solid oxide electrolysers. The startup behaviour of the electrolyser will be improved by considering the relation of the startup cost to the fluctuating electricity price. This provides a more accurate startup cost than currently used in the literature, where the startup cost is considered fixed \cite{baumhof_92optimization_2023, matute_120multi-state_2021, varela_122modeling_2021, zheng_optimal_2022}.

Furthermore, the models are adapted to allow for external heat to be used in the electrolysis process.  For each electrolyser type, the model is modified to make use of heat as applicable. The modelling contribution of this paper is therefore twofold; the startup behaviour is modelled more accurately, and heat addition is integrated into the models. The models are applied to a case study to discuss the results and implications of the improvements made.

%The remainder of the paper is structured as follows. In Section \ref{sec:integrating_heat}, heat integration in electrolysers is discussed. In Section \ref{sec:models} the electrolyser models with heat integration are formulated. Section \ref{sec:casestudy} covers the case study on a 15 MW PEM and SOE electrolyser. Finally, conclusions are drawn.
%%%%%%%%%%%%%%%%%%%%%%%%%%%%%%%%%%%%%%%%
%   Literature Review 
%%%%%%%%%%%%%%%%%%%%%%%%%%%%%%%%%%%%%%%%
\section{External heat integration}
\label{sec:integrating_heat}
External heat integration in electrolysis technologies—such as Alkaline, Proton Exchange Membrane (PEM), and Solid Oxide Electrolysers (SOE)—can significantly enhance their performance and efficiency. SOEs, in particular, benefit greatly from external heat due to their high operating temperatures (800–1000$\degree$C), which align well with the availability of high-temperature heat sources \cite{petipas_72benefits_2014}. The integration of heat in SOEs can reduce the overall electrical energy requirement by supplying part of the necessary thermal energy directly through external heat sources or regeneration processes. This approach allows the electrolysis process to operate below the thermoneutral voltage while keeping the operating temperature constant, reducing the electrical input needed for water electrolysis. Various auxillary processes, such as preheating during cold start and heating during standby, are able to use external heat. An overview of the heat integration possibilities is provided in Table \ref{tab:combinationsnuclearelectrolyser}.
 \begin{table}[tb]
    \centering
    \caption{Use of low- and high-temperature heat for different electrolysers}
    \begin{tabular}{|p{1cm}|p{2.5cm}|p{4cm}|}
        \cline{2-3} % Horizontal line from column 2 to 3
        \multicolumn{1}{c|}{} & PEM (20-100$\degree C$)  \& Alkaline (40-90$\degree C$) & \multicolumn{1}{c|}{SOE (800-1000$\degree C$)} \\
        \hline
        \begin{flushleft} 
        \small {Low-T heat (280-325$\degree C$)}
        \end{flushleft} &  & \begin{itemize}[leftmargin=*]
            \item Heating and evaporating inlet water
            \item Heating electrolyser during cold start, in combination with an electrical heater 
        \end{itemize}\\
        
       \cline{1-1} \cline{3-3}
        \begin{flushleft}
            \small High-T heat (750-1000$\degree C$)
        \end{flushleft} & \vspace{-20mm}\begin{itemize}[leftmargin=*]
            \item Heating inlet water
            \item Keeping temperature constant in standby
            \item Heating electrolyser during cold start
        \end{itemize}
         &
        \begin{itemize}[leftmargin=*]
            \item Heating and evaporating inlet water
            \item Heating steam  from pinch point to operating temperature
            \item Keeping temperature constant in standby
            \item Heating electrolyser during cold start
            \item Heating sweep gases to supply heat directly to the electrolyser
        \end{itemize} \\
        \hline
    \end{tabular}
    \label{tab:combinationsnuclearelectrolyser}    
\end{table}

\cite{petipas_72benefits_2014} explored the effects of using external heat sources in the SOE operating range, highlighting that high-temperature heat can be employed to preheat inlet water and superheat steam using regenerative heat exchangers. This configuration obtains a system efficiency improvement of 18\% by minimizing the amount of electrical energy required to achieve the high operational temperatures of the SOE. \cite{zhang_75efficiency_2012} evaluates the efficiency of high-temperature steam electrolytic systems coupled with various types of nuclear reactors. It specifically examines how heat from different nuclear reactor designs can be used to generate the steam needed for the electrolysis process, aiming to optimize the overall system performance.

In contrast, the Alkaline and PEM electrolysers operate at significantly lower temperatures (20–100$\degree$C for PEM and 40–90$\degree$ C for Alkaline), and cannot directly use external heat to drive the electrolysis reaction due to their inherently exothermal operating regimes. However, heat can still be beneficial for auxillary processes, such as preheating the inlet water or maintaining the electrolyser temperature during standby modes, thereby reducing the need for electrical heating and enhancing overall system efficiency.

%%%%%%%%%%%%%%%%%%%%%%%%%%%%%%
%      METHODS
%%%%%%%%%%%%%%%%%%%%%%%%%%%%%%
\section{Operational models}
\label{sec:models}

\subsection{Linearisation of electrochemical models}
\begin{figure}
    \begin{subfigure}{.49\linewidth}
      \centering
      \includegraphics[width=\textwidth]{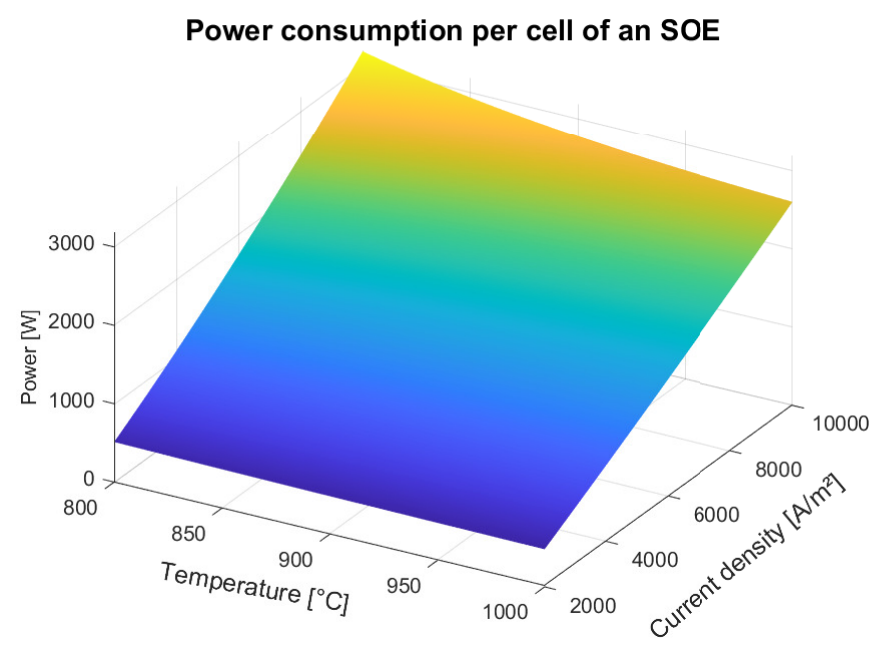}
      \caption{}
    \end{subfigure}%
    \begin{subfigure}{.49\linewidth}
      \centering
      \includegraphics[width=\linewidth]{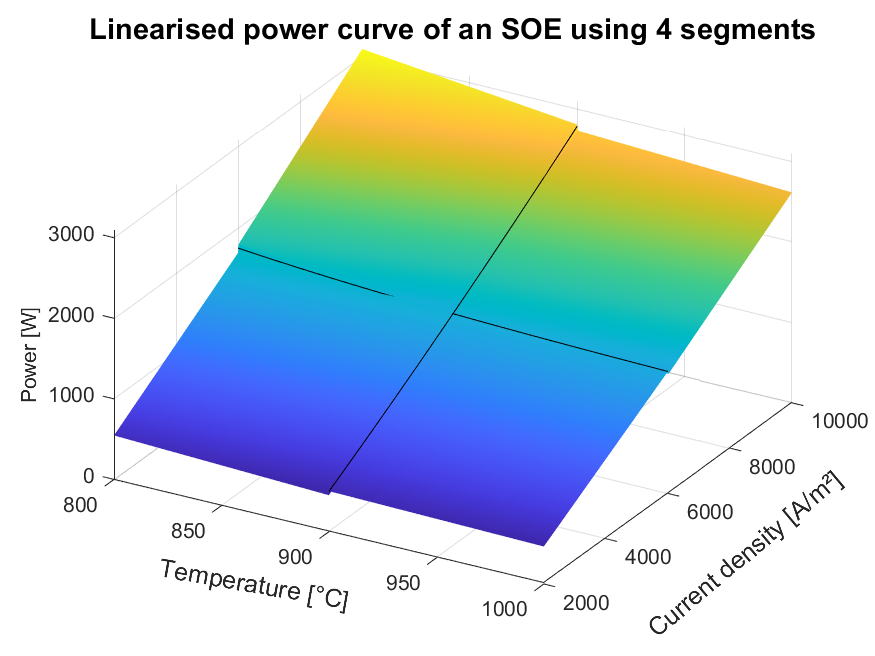}
      \caption{}
      \label{fig:linpowerSOEC}
    \end{subfigure}
\caption{(a) Power consumed per cell by a SOE as a function of temperature and current density based on the underlying electrochemical effects described in Appendix A and (b) its linearization with four segments}
\label{fig:powerSOEC}
\end{figure}

A piecewise linearisation is applied to the power curves of the PEM and SOE electrolyser. An example of such a nonconvex power curve of the SOE is shown in Figure \ref{fig:powerSOEC} together with its division into multiple segments. Each segment approximates the power values in its own segment by minimising the square of linearisation error.
The least squares method is used to determine the coefficients \(a\), \(b\), and \(c\) for the best-fitting plane \(P = a \cdot T + b \cdot j + c\) in each segment, where \(T\) represents temperature and \(j\) current density. For linear regression, only power values at the minimum and maximum current densities of each segment are used across the temperature range. This approach ensures linearisation is exact at the segment boundaries and avoids significant over- or underestimation at the edges, which could cause discontinuities between segments.

With one segment, the mean relative error is \(4.83\%\), reducing to \(1.14\%\) with four segments and \(0.50\%\) with nine. %The maximum local relative error, observed at low current densities, decreases from \(28.39\%\) for one segment to \(7.53\%\) for four and \(3.32\%\) for nine. 
Applying the same method to the PEM electrolyser, the mean relative errors for one, four, and nine segments are \(13.44\%\), \(3.32\%\), and \(1.51\%\), respectively. These are higher than for the SOE due to the PEM electrolyser's larger operating current density range. %The maximum relative errors are \(114.31\%\), \(22.75\%\), and \(8.70\%\) for one, four, and nine segments, respectively.

\subsection{Solid oxide electrolyser operational model}
The SOE model considers only the standby and production states, excluding the off-state due to the long cold start duration of up to 6.5 hours and the risk of thermal cycling damage associated with frequent cold starts \cite{irena_128green_2020}. As such, cold starts are to be avoided and omitted from the operational optimisation.\footnote{Table \ref{tab:resultsSOE} shows that the number of hours in production state is high, confirming this assumption.}

Drawing partly on the alkaline electrolyser model described by \cite{zheng_optimal_2022}, the SOE model is formulated as follows. The objective of the operational model is to maximise profit \(P\), as in Eq.~\eqref{eq:SOEC1obj}, with $x = \{I_t, \dot{Q}^{heat}_t, \qstandby, \qcool, p^b_t, s^b_t, S^b_{t,m,n}, \delta^1_t, T_t, \pcell\}
$. Electricity prices are exogenous and vary throughout the day, while a constant hydrogen price is assumed. The decision variables include the current \(I\) \([A]\), the heat added during production \(\dot{Q}^{heat}_t\) \([W]\), heat added in standby \(\qstandby\) \([W]\), cooling supplied \(\qcool\) \([W]\), binary variables \(p^b_t\) and \(s^b_t\) indicating the production and standby states, the segment binary variable \(S^b_{t,m,n}\) for linearisation, and the slack variable \(\delta^1_t\). Temperature \(T_t\) \([K]\) and electrolyser power \(\pcell\) \([W]\) depend on current and temperature at prior or current timesteps.

To simplify notation, \(\ptotal\), \(\dot{m}^H_t\) \([kg/h]\), and \(\qwater\) are introduced as expressions of the decision variables. The model Eqs. are formulated using a power curve approximation divided into \(M\) sections for temperature and \(N\) sections for current density, resulting in \(M \cdot N\) segments. The Eqs. assume low-temperature heat supply to the SOE, with \(\Delta t\) representing the timestep duration in hours.\footnote{In this study, 15-minute intervals are modelled, setting \(\Delta t = 0.25\).}
\begin{equation}
    \mathop{Max}_x P = \sum_{t \in\setT} ( \dot{m}^H_t \cdot \lambda^H  - \ptotal \cdot \lambda^E_t ) \cdot \Delta t
    \label{eq:SOEC1obj}
\end{equation}
Eq.\eqref{eq:SOEC2pE} describes the electrical power required without heat integration. It includes the power per cell, $\pcell$ [W], the number of cells $n_c$, the total heat demand, the efficiency of the electrical heater $\eta_{EH}$, the compressor power $\pcomp$ [W], and the cooling power\footnote{\cite{petipas_72benefits_2014} approximate the cooling power $p^{cool}_t$ as $p^{cool}_t = \frac{\qcool}{400}$, where $\qcool$ is the thermal cooling power}. The low-temperature heat demand is used exclusively to heat water to $100^\circ C$ and evaporate it, represented by $\qwater = (c_p \Delta T_a + \Delta H)\dot{m}^{H_2O}_t$

Eq.\eqref{eq:SOEC3pE} extends this formulation to include heat integration, where the heat cost is modelled as the opportunity cost of not converting the heat into electricity via a steam turbine. Most heating to operating temperature is achieved through regeneration. The final 40°C increase to reach operating temperature is applied by $\dot{Q}^{steam}_t = 40 \degree C \cdot c_p\cdot \dot{m}^{H_2O}_t$
\begin{align}
    \ptotal = n_c\cdot \pcell + \frac{\qstandby + \dot{Q}^{steam}_t + \dot{Q}^{heat}_t}{\eta_{EH}} \nonumber \\
    +  \qwater\cdot \eta_{ST}+ \pcomp + p^{cool}_t   \hspace{0.5cm} \forall t \in \setT
    \label{eq:SOEC2pE}
\end{align}
\begin{align}
    \ptotal = n_c\cdot \pcell + (\qstandby + \qwater + \dot{Q}^{steam}_t \nonumber \\ + \dot{Q}^{heat}_t) \cdot \eta_{ST} 
    + \pcomp + p^{cool}_t   \hspace{0.5cm} \forall t \in \setT
    \label{eq:SOEC3pE}
\end{align}
The hydrogen production rate is expressed using Eq. \eqref{eq:SOEC1mdot}. The water mass flow rate in Eq. \eqref{eq:SOEC1mdotH2O} follows from that. The current and temperature determine the electrical power consumption using Eq. \eqref{eq:SOEC1power}. Here $a_{m,n}$, $b_{m,n}$ and $c_{m,n}$ are the coefficients of the plane that approximates the power curve in the $(m,n)$ segment, which is located in the $m^{th}$ current density section and the $n^{th}$ temperature section. 
\begin{equation}
     \dot{m}^H_t = n_c M_{H_2} \frac{I_t}{2 F} \hspace{0.5cm} \forall t \in \setT
    \label{eq:SOEC1mdot}
\end{equation}
\begin{equation}
    \dot{m}^{H_2O}_t = \frac{M_{H_2O}}{M_{H_2}}\dot{m}^H_t\quad \forall t \in \setT
    \label{eq:SOEC1mdotH2O}
\end{equation}
\begin{equation}
    \pcell = \sum_{m \in M, n \in N } S^{b}_{t,m,n}(a_{m,n}T_t + b_{m,n}\frac{I_t}{A} + c_{m,n}) + \delta^1_{t} \quad \forall t \in \setT
    \label{eq:SOEC1power}
\end{equation}
The model is subject to Eq. \eqref{eq:SOEC1sumsegments}-\eqref{eq:SOEC1sumbinary}. Constraints \eqref{eq:SOEC1sumsegments}, \eqref{eq:SOEC1jsegments1}, and \eqref{eq:SOEC1Tsegments1} ensure that the binary variable representing the active segment is set to one. Eqs. \eqref{eq:SOEC1Iborder} and \eqref{eq:SOEC1Tborder} enforce current and temperature limits, while Eq. \eqref{eq:SOEC1pborders} ensures the power is zero when not in the production state. The slack variable associated with power is constrained to zero unless in standby mode by Eq. \eqref{eq:SOEC1delta1}. Eq.\eqref{eq:SOEC1ps} governs the heating demand during standby. Compressor power is defined in Eq. \eqref{eq:SOEC1pc}, where $\alpha$ depends on the pressure ratio and compressor efficiency. Eqs. \eqref{eq:SOEC1coolingborders} and \eqref{eq:SOEC1heatingborders} set the cooling and direct heat addition variables to zero when not in the production state. Eq. \eqref{eq:SOEC1temp} expresses the temperature as a function of the previous timestep, based on losses and the heat flux $\dot{Q}_{t}  = \dot{Q}^{standby}_{t}  + \dot{Q}^{heat}_{t} - \dot{Q}^{cool}_{t}$. Eq. \eqref{eq:SOEC1sumbinary} ensures that only one binary state variable is active at any time.
\begin{equation}
     \sum_{m \in M, n \in N } S^{b}_{t,m,n} = 1 \quad \forall t \in \setT
    \label{eq:SOEC1sumsegments}
\end{equation}
\begin{equation}
     \sum_{m \in M, n \in N } S^{b}_{t,m,n} \underline{j_m} \leq \frac{I_t}{A} \leq  \sum_{m \in M, n \in N } S^{b}_{t,m,n} \overline{j_m} \quad \forall t \in \setT
    \label{eq:SOEC1jsegments1}
\end{equation}
\begin{equation}
     \sum_{m \in M, n \in N } S^{b}_{t,m,n}  \underline{T_n} \leq T_t \leq \sum_{m \in M, n \in N } S^{b}_{t,m,n}  \overline{T_n} \quad \forall t \in \setT
    \label{eq:SOEC1Tsegments1}
\end{equation}
\begin{equation}
    p^b_t \underline{j}A \leq I_t \leq p^b_t \overline{j}A \hspace{0.5cm} \forall t \in \setT
    \label{eq:SOEC1Iborder}
\end{equation}
\begin{equation}
     \underline{T} \leq T_t \leq \overline{T} \quad \forall t\in \setT
    \label{eq:SOEC1Tborder}
\end{equation}
\begin{equation}
    \frac{T_t - T_a}{R_t}s^b_t \leq \qstandby\leq Ms^b_t\quad \forall t \in \setT
    \label{eq:SOEC1ps}
\end{equation}
\begin{equation}
    0 \leq \pcell \leq p^b_t\overline{p}  \hspace{0.5cm} \forall t \in \setT
    \label{eq:SOEC1pborders}
\end{equation}
\begin{equation}
   p^c_{t} =  \alpha \dot{m}^{H}_t \quad \forall t \in \setT
    \label{eq:SOEC1pc}
\end{equation}
\begin{align}
     \frac{C_h(T_t - T_{t-1})}{\Delta t} = n_c(p^{cell}_{t-1}  - U_{tn}I_{t-1}) \nonumber \\ - \frac{T_{t-1}- T_a}{R_t} + \dot{Q}_{t-1} \quad \forall t \in \setTzero
    \label{eq:SOEC1temp}
\end{align}
\begin{equation}
    0 \leq \dot{Q}^{cool}_{t} \leq p^b_tM\quad \forall t \in \setT
    \label{eq:SOEC1coolingborders}
\end{equation}
% DIFFERENT FROM PEM
\begin{equation}
    0 \leq \dot{Q}^{heat}_{t} \leq p^b_tM\quad \forall t \in \setT
    \label{eq:SOEC1heatingborders}
\end{equation}
\begin{equation}
    -Ms^b_t \leq \delta^1_{t} \leq Ms^b_t\quad \forall t \in \setT 
    \label{eq:SOEC1delta1}
\end{equation}
\begin{equation}
    p^b_t + s^b_t = 1\quad \forall t \in \setT
    \label{eq:SOEC1sumbinary}
\end{equation}
\subsection{PEM electrolyser operational model}
The PEM electrolyser model includes on, off, and standby states, maximising profit \( P \) as in Eq. \eqref{eq:PEM1obj}, with $x = \{ I_t, \qstandby, \qcool, p^b_t, s^b_t, i^b_t, Z^b_t, S^b_{t,m,n}, \delta^1_t, T_t, \pcell \}
$. Cold start costs, \( C^{CS}_t \), account for reduced hydrogen output and electricity use during start-up (Eq. \eqref{eq:PEM1CCS}) and are linked to electricity and hydrogen prices for greater accuracy. Cold starts require no external heat and are assumed to last 10 minutes, while hot starts are instantaneous and cost-free, reflecting the PEM's operational flexibility.
\begin{equation}
    \mathop{Max}_x P = \sum_{t \in \setT} ( \dot{m}^H_t \cdot \lambda^H  - \ptotal \cdot \lambda^E_t ) \cdot \Delta t - \sum_{t \in \setTzero } C^{CS}_t Z^b_t
    \label{eq:PEM1obj}
\end{equation}
\begin{equation}
    C^{CS}_t = \Delta t^{CS} \cdot (\dot{m}^H_{\underline{T}, \underline{I}} \cdot \lambda^H - p^E_{\underline{T}, \underline{I}} \cdot \lambda^E_t) \hspace{1cm} \forall t \in \setT
    \label{eq:PEM1CCS}
\end{equation}
Eq. \eqref{eq:PEM1pE} calculates the total electrical power required without heat integration, while Eq. \eqref{eq:PEM2pE} includes heat integration. Heating inlet water from ambient temperature to \( 100\degree \text{C} \) is partially achieved through regeneration.\footnote{We determined that the outlet hydrogen stream at \( 100\degree \text{C} \) can heat inlet water from \( 20\degree \text{C} \) to \( 46.8\degree \text{C} \), leaving \( \Delta T^{water} = 53.2\degree \text{C} \); details in \cite{derez_improvement_2023}} The remaining heat demand, \( \qwater \), is described by Eq. \eqref{eq:PEM1Qwater}, which assumes operation at maximum temperature and efficiency. This assumption is justified as the PEM electrolyser typically operates under these conditions, and the heat demand is minor compared to other energy requirements. Heat is used to warm inlet water and supply standby thermal power, with costs representing the opportunity cost of forgone electricity generation.
\begin{equation}
    \ptotal = n_c\cdot \pcell + \frac{\qstandby + \qwater}{\eta_{EH}} + \pcomp + p^{cool}_t  \hspace{0.5cm} \forall t \in \setT
    \label{eq:PEM1pE}
\end{equation}
\begin{equation}
    \ptotal = n_c\cdot \pcell + (\qstandby + \qwater)\cdot \eta_{ST} + \pcomp + p^{cool}_t  \hspace{0.5cm} \forall t \in \setT
    \label{eq:PEM2pE}
\end{equation}
\begin{equation}
   \dot{Q}^{water}_{t} =  \dot{m}^{H_2O}_tc^{water}_p\cdot \Delta T^{water} \hspace{1cm} \forall t \in \setT
    \label{eq:PEM1Qwater}
\end{equation}
The model is subject to constraints \eqref{eq:SOEC1sumsegments}-\eqref{eq:SOEC1coolingborders} and \eqref{eq:PEM1QCS}-\eqref{eq:PEM1binarystartup2}. In Eq. \eqref{eq:SOEC1temp} the heat flux is given by $\dot{Q}_t = \dot{Q}^{standby}_t -\dot{Q}^{CS}_{t} - \dot{Q}^{cool}_{t}$. With $\dot{Q}^{CS}_t$, as described in Eq. \eqref{eq:PEM1QCS}. Eqs. \eqref{eq:PEM1delta1}-\eqref{eq:PEM1binarystartup2} extend the binary logic to account for the three operating states.
%THESE ARE CHANGED
\begin{equation}
    \dot{Q}^{CS}_t = \frac{\Delta t^{CS}}{\Delta t} (n_c p_{\underline{T}, \underline{I}} - n_cU_{tn}\underline{I}) \hspace{1cm} \forall t \in \setT
    \label{eq:PEM1QCS}
\end{equation}
%THESE ARE ADDED BINARY LOGIC
\begin{equation}
    -M(s^b_t+i^b_t) \leq \delta^1_{t} \leq M(s^b_t+i^b_t)\hspace{1cm} \forall t \in \setT 
    \label{eq:PEM1delta1}
\end{equation}
\begin{equation}
    p^b_t + s^b_t + i^b_t = 1\hspace{1cm} \forall t \in \setT
    \label{eq:PEM1sumbinary}
\end{equation}
% \begin{equation}\small
%     p^b_t = p^b_0 \hspace{1cm} t = t_1
%     \label{eq:PEM1pbo}
% \end{equation}
% \begin{equation}\small
%     s^b_t = s^b_0 \hspace{1cm} t = t_1
%     \label{eq:PEM1sbo}
% \end{equation}
% \begin{equation}
%     i^b_t = i^b_0 \hspace{1cm} t = t_1
%     \label{eq:PEM1ibo}
% \end{equation}
\begin{equation}
    s^b_t + i^b_{t-1} \leq  1\hspace{1cm} \forall t \in \setTzero
    \label{eq:PEM1offtostandby}
\end{equation}
\begin{equation}
    s^b_{t-1} + i^b_{t} \leq  1\hspace{1cm} \forall t \in \setTzero
    \label{eq:PEM1standbytooff}
\end{equation}
\begin{equation}
    i^b_{t-1} + p^b_{t} -1 \leq Z^b_{t} \leq i^b_{t-1}\hspace{1cm} \forall t\in \setTzero
    \label{eq:PEM1binarystartup1}
\end{equation}
\begin{equation}
    Z^b_{t} \leq p^b_{t}\hspace{1cm} \forall t\in \setTzero
    \label{eq:PEM1binarystartup2}
\end{equation}
%%%%%%%%%%%%%%%%%%%%%%%%%%%%%%
%      CASE STUDY
%%%%%%%%%%%%%%%%%%%%%%%%%%%%%%
\section{Case study}
\label{sec:casestudy}
\subsection{Data \& cases}
This section applies the models from Section \ref{sec:models} to a case study. The linearised models were implemented in Julia and solved using Gurobi.\footnote{The Julia implementation is publicly available at: \ \url{https://github.com/Matthias-Derez/Electrolyser-Models}} Electrolyser parameters are scaled to consume 15 MW of electrical power at maximum temperature and current density (see \ref{app:parameters}). To assess profitability and the impact of heat integration, daily profits are calculated for each day of a full year using 2019 Belgian day-ahead electricity prices, avoiding distortions from COVID-19 and the energy crisis. Hydrogen price levels range from 2.5 to 5.5 EUR/kg.

A rolling horizon approach optimises daily operation independently. The final operational state and temperature from each day serve as starting conditions for the next, ensuring realistic transitions and startup costs.\footnote{Validation was performed by comparing rolling horizon results to full-week optimisations for randomly selected weeks.} 

We consider five cases: 1) The case \textit{ SOE without heat integration} serves as a reference for the SOE model in which no heat integration is applied in any form. Hence, Eq. \eqref{eq:SOEC2pE} applies, and $\dot{Q}^{heat}_t$ is constraint to zero. 2) In the case \textit{SOE with low-T heat integration} external heat is used in standby and during start-up. So, Eq. \eqref{eq:SOEC3pE} applies, but $\dot{Q}^{heat}_t$ is constraint to zero. 3) In the case \textit{SOE with high-T integration}, direct addition of external heat is considered, which reduces the electrical power required. Hence, Eq. \eqref{eq:SOEC3pE} applies, and $\dot{Q}^{heat}_t$ is nonzero. 4) The case \textit{PEM without heat integration} serves as a reference for the PEM model in which no heat integration is applied in any form. Hence, Eq. \eqref{eq:PEM1pE} applies. 5) In the case \textit{PEM with heat integration}, external heat is used in standby and during start-up. Therefore, Eq. \eqref{eq:PEM2pE} applies.
\subsection{Results heat integration in a SOE}
\begin{figure*}
 \centering
  \begin{subfigure}{0.45\textwidth}
  \includegraphics[width=\linewidth]{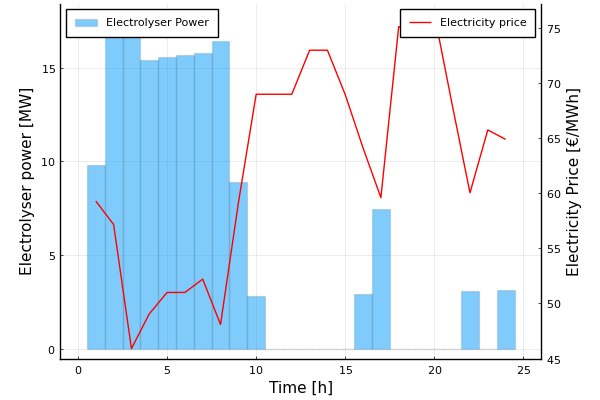}
  \caption{}
  \label{fig:SOECpowerprice}
  \end{subfigure}
  \begin{subfigure}{0.45\textwidth}
  \includegraphics[width=\linewidth]{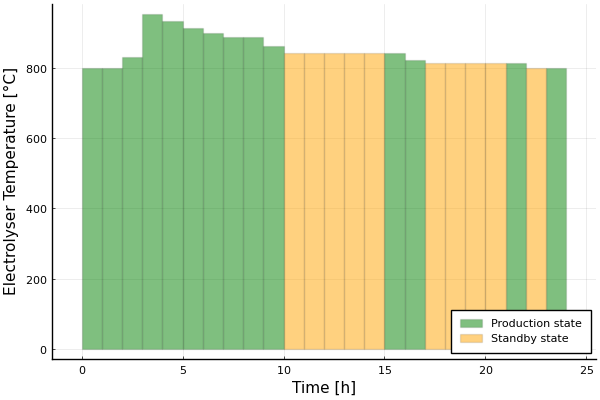}
  \caption{}
  \label{fig:SOECtempbarscolours}
  \end{subfigure}
  \caption{(a) SOE power and electricity price, and (b) temperature and operational state, for 5th Jan 2019 using 2.5 EUR/kg $\si{H_2}$}
\end{figure*}
Figures \ref{fig:SOECpowerprice} and \ref{fig:SOECtempbarscolours} illustrate the typical operation of the SOE model without heat integration over one day. Figure \ref{fig:SOECpowerprice} depicts day-ahead electricity prices and electrolyser power, while Figure \ref{fig:SOECtempbarscolours} shows temperature evolution, with bar colours indicating operational states. During periods of high electricity prices, the electrolyser enters standby. At the thermoneutral voltage temperature of $900 \degree C$ (achieved at a current density of 10,000 A/m\(^2\)), no external heating is needed. Operating above this temperature increases efficiency but reduces profit due to higher heating costs, while operating below it reduces efficiency and is avoided unless standby is anticipated.  

Integrating heat significantly improves profitability. Low-temperature heat increases profits by \(17.1\%-4.0\%\) for hydrogen prices of 2.5–5.5 EUR/kg, while high-temperature heat further increases profits by \(23.3\%-5.4\%\). Table \ref{tab:improvementheat} summarises these relative improvements.  

The system efficiency increases from \(77.77\%\) without heat integration to \(85.63\%\) with low-temperature heat and \(88.73\%\) with high-temperature heat. These values align with the literature: efficiencies of \(76-81\%\) (LHV basis) without heat integration \cite{kecebas_62electrochemical_2019} and \(89.5\%\) with high-temperature heat \cite{zhao_84system_2021}. Efficiency remains constant across hydrogen prices, as even at 2.5 EUR/kg, marginal production costs are covered for most hours.  

Profitability was assessed by comparing profits with investment costs. Based on \cite{irena_128green_2020}, a 15 MW SOE costs 34.5 MEUR. Assuming a 20,000-hour lifetime and a \(3\%\) discount rate, break-even analysis reveals that average daily profits are generally insufficient to cover investment costs, due to high capital costs and limited lifetime. Only at 5.5 EUR/kg hydrogen, with low-cost heat integration, is profitability sufficient to break even. Detailed results are presented in Table \ref{tab:resultsSOE} in Appendix C.

\subsection{Results heat integration in a PEM}
Figures \ref{fig:PEMpowerprice} and \ref{fig:PEMtempbarscolours} show an example of typical PEM electrolyser behaviour, using the \textit{PEM without heat integration} model. Because water electrolysis is an exotherm process, the temperature is close to the maximum temperature during operation. Furthermore, the off-state is preferred over the standby state, as the temperature can increase rapidly after a cold start.
\begin{table}[tb]
    \centering
    \begin{tabular}{cccc}
        \hline
        \hline
        Hydrogen   & SOE with  & SOE with & PEM with  \\
        Price & low-T heat & high-T heat & low/high-T heat \\
        (EUR/kg) & (\%) & (\%) & (\%) \\
        \midrule
        2.5 & 17 & 23 & 1.8 \\
        3.5 & 8 & 12 & 1 \\
        4.5 & 6 & 8 & 0.6 \\
        5.5 & 4 & 6 & 0.4 \\
        \hline
        \hline
    \end{tabular}
        \caption{Relative improvement in average profit per day for a SOE and PEM for different degrees of heat integration, compared to no heat integration}
    \label{tab:improvementheat}
\end{table}
The impact of an external low-cost heat source on profitability is modest, increasing average daily profits by just \(1.83\%-0.44\%\), as shown in Table \ref{tab:improvementheat}. This is because the off-state is generally preferred over standby, and the heat required to bring water to operating temperature is minimal compared to other energy demands.  

For PEM electrolysers, power loading significantly influences efficiency, making partial loading a viable strategy when operating near marginal costs. This trade-off arises from the wide current density range, which leads to substantial efficiency variations. For instance, at 1500 A/m\(^2\), the hydrogen production rate is low but efficient, resulting in low marginal costs. Conversely, at 20,000 A/m\(^2\), production is high but less efficient, with higher marginal costs. This efficiency-cost balance can lead to suboptimal results and relatively low daily profits depending on electricity and hydrogen prices.  

Compared to SOE, the lower capital costs and longer lifetimes of PEM electrolysers make them more economically viable in certain scenarios. Assuming an 80,000-hour lifetime, average daily profits are sufficient to cover investment costs at a hydrogen price of 3.5 EUR/kg. For a 50,000-hour lifetime, this threshold rises to 4.5 EUR/kg . These calculations use a CAPEX of 900 EUR/kW, based on \cite{irena_128green_2020}.\footnote{The numerical results are contingent on the specific year of electricity price data and the nominal power of the electrolyser.}
\section{Conclusion}
This study develops operational models for SOE and PEM electrolysers, incorporating electrochemical equations that relate power consumption to temperature and current density. The models are enhanced to integrate external heat sources, using a piecewise-linear approximation of the power curve. Startup costs are refined to better reflect real electrolyser behaviour in response to instantaneous electricity prices.

We examine key characteristics of these technologies, focusing on the impact of temperature and direct heat supply on efficiency. Increasing temperature reduces irreversible losses, improving efficiency, while direct heat integration allows SOE to reduce electrical power consumption, boosting overall efficiency. PEM electrolysers cannot directly use external heat for operation but can benefit from heat to maintain standby mode, a feature both technologies can utilise.

A case study of a 15 MW PEM and SOE estimates profit increases from heat integration. The SOE's daily profits rise by up to 23.3\%, while PEM profits increase by just 1.83\%. However, despite heat integration, the SOE requires a hydrogen price of 5.5 EUR/kg to break even, and PEM electrolyser costs remain lower due to a more favourable CAPEX and longer lifetime. Heat integration does not significantly reduce the break-even price for PEM.

The models developed in this paper can be used to plan the optimal dispatch of the different electrolyser types. However, because of their efficient formulation, the models can also be used in energy system models, e.g., for planning studies, where  a constant electrolyser efficiency is often assumed. The models developed in this study take into account the varying electrolyser efficiency and increase the accuracy of the results. %In future research, the model for the solid oxide electrolyser could be adapted to consider the off-state as well. In this way, the effect of heat integration on cold start costs can also be taken into account.
\label{sec:conclusion}

%\section*{Acknowledgment}

%The preferred spelling of the word ``acknowledgment'' in America is without an ``e'' after the ``g''. Avoid the stilted expression ``one of us (R. B. G.) thanks $\ldots$''. Instead, try ``R. B. G. thanks$\ldots$''. Put sponsor acknowledgments in the unnumbered footnote on the first page.

\bibliographystyle{IEEEtran}
\bibliography{cas-refs}

\appendix
\input{appendix-A}

\input{appendix-B}

\input{appendix-C}

%\begin{thebibliography}{00}

%\end{thebibliography}

\end{document}

%% file: math.tex
%%% Math Symbols

% Sets
\newcommand{\setD}{\mathcal{D}}
\newcommand{\setH}{\mathcal{H}}
\newcommand{\setM}{\mathcal{M}}
\newcommand{\setP}{\mathcal{P}}
\newcommand{\setR}{\mathcal{R}}
\newcommand{\setY}{\mathcal{Y}}
\newcommand{\setTY}{\mathcal{Y}_T}
\newcommand{\setYzero}{\setY^0}
\newcommand{\setF}{\mathcal{F}}
\newcommand{\setQ}{\mathcal{Q}}
\newcommand{\setQs}{\mathcal{Q}^S}
\newcommand{\setT}{\mathcal{T}}
\newcommand{\setTzero}{\mathcal{T}_0}

% Variables 
\newcommand{\bc}{b^{\mathrm{C}}_{y,p}}
\newcommand{\bci}{b^{\mathrm{C,i}}_{y,p}}
\newcommand{\bcstar}{b^{\mathrm{C}}_{y^{*},p}}
\newcommand{\bi}{b^{\mathrm{I}}_{y}}
\newcommand{\bii}{b^{\mathrm{I,i}}_{y}}
\newcommand{\bistar}{b^{\mathrm{I}}_{y^{*}}}
\newcommand{\cmsr}{c^{\mathrm{MSR}}_{y,m}}
\newcommand{\cp}{cp^{\mathrm{C}}_{y,r}}
\newcommand{\cpstar}{cp^{\mathrm{C}}_{y^{*},r}}
\newcommand{\cpr}{cp^{\mathrm{R}}_{y,r}}
\newcommand{\cprstar}{cp^{\mathrm{R}}_{y^{*},r}}
\newcommand{\ei}{e^{\mathrm{I}}_{y}}
\newcommand{\eistar}{e^{\mathrm{I}}_{y^{*}}}
\newcommand{\gc}{g^{\mathrm{C}}_{y,d,h,p}}
\newcommand{\gcstar}{g^{\mathrm{C}}_{y^{*},d,h,p}}
\newcommand{\gr}{g^{\mathrm{R}}_{y,d,h,r}}
\newcommand{\grn}{g^{\mathrm{R,NB}}_{y,r}}
\newcommand{\pets}{\lambda^{\mathrm{ETS}}_{y}}
\newcommand{\petsi}{\lambda^{\mathrm{ETS,i}}_{y}}
\newcommand{\peom}{\lambda^{\mathrm{EOM}}_{y,d,h}}
\newcommand{\pres}{\lambda^{\mathrm{REC}}_{y}}
\newcommand{\pfp}{\lambda^{\mathrm{FP}}}
\newcommand{\pcfd}{\lambda^{\mathrm{CfD}}}
\newcommand{\pinv}{\lambda^{\mathrm{INV}}}
\newcommand{\pcg}{\lambda^{\mathrm{CG}}}
\newcommand{\msr}{msr_{y,m}}
\newcommand{\preseom}{R^{\mathrm{EOM,i}}}
\newcommand{\presets}{R^{\mathrm{ETS,i}}}
\newcommand{\presrec}{R^{\mathrm{REC,i}}}
\newcommand{\dresc}{R^{\mathrm{C,i}}_p}
\newcommand{\dresr}{R^{\mathrm{R,i}}_r}
\newcommand{\dresi}{R^{\mathrm{I,i}}}
\newcommand{\supply}{S_{y}}
\newcommand{\tnac}{tnac_{y}}
\newcommand{\xmsr}{x^{\mathrm{MSR}}_{y,m}}
% Parameters
\newcommand{\deltay}{\delta_y}
\newcommand{\Ay}{A_y}
\newcommand{\Aysp}{A^{\mathrm{SP}}_{y}}
\newcommand{\av}{AV_{d,h,r}}
\newcommand{\ci}{CI_{p}}
\newcommand{\legcapc}{\overline{CP_{y,p}}}
\newcommand{\legcapr}{\overline{CP_{y,r}}}
\newcommand{\dem}{D_{y,d,h}}
\newcommand{\macc}{\mathcal{F}_y(\pets)}
\newcommand{\icc}{IC^\mathrm{C}_{p}}
\newcommand{\icr}{IC^\mathrm{R}_{r}}
\newcommand{\ltc}{LT^\mathrm{C}_{y,y^{*},p}}
\newcommand{\ltr}{LT^\mathrm{R}_{y,y^{*},r}}
\newcommand{\neom}{N^{\mathrm{EOM}}}
\newcommand{\nets}{N^{\mathrm{ETS}}}
\newcommand{\restarget}{RT_{y}}
\newcommand{\rhoETS}{\rho^{\mathrm{ETS}}}
\newcommand{\emcap}{\overline{S_y}}
\newcommand{\svc}{SV^\mathrm{C}_{y,p}}
\newcommand{\svr}{SV^\mathrm{R}_{y,r}}
\newcommand{\vc}{VC_{p}}
\newcommand{\weight}{W_{d}}
\newcommand{\llxmsr}{\underline{\xmsr}}
\newcommand{\ulxmsr}{\overline{\xmsr}}
% Abbreviations
\newcommand{\CO}{CO$_{\mathrm{2}}$ }
\newcommand{\tCO}{tCO$_{\mathrm{2}}$ }
% Operations
\newcommand{\sumy}{\sum_{y \in \setY}}
\newcommand{\sumystarx}{\sum_{y^{*}=1}^{y}}
\newcommand{\sumyt}{\sum_{y* \in \setY}}
\newcommand{\sumd}{\sum_{d \in \setD}}
\newcommand{\sumh}{\sum_{h \in \setH}}
\newcommand{\sump}{\sum_{p \in \setP}}
\newcommand{\sumr}{\sum_{r \in \setR}}
\newcommand{\sumq}{\sum_{q \in \setQ}}
\newcommand{\sumf}{\sum_{f \in \setF}}
\newcommand{\ally}{\forall y \in \setY}
\newcommand{\allyzero}{\forall y \in \setY^0}
\newcommand{\alld}{\forall d \in \setD}
\newcommand{\allh}{\forall h \in \setH}
\newcommand{\allp}{\forall p \in \setP}
\newcommand{\allr}{\forall r \in \setR}
\newcommand{\allt}{\forall y \in \setTY}
\newcommand{\allf}{\forall f \in \setF}
%%% New hydrogen related variables \& parameters
\newcommand{\setS}{\mathcal{S}}
\newcommand{\ph}{\lambda^{\mathrm{H}}_{y,d}}
\newcommand{\gh}{g^{\mathrm{H}}_{y,d,h,q}}

\newcommand{\ghstar}{g^{\mathrm{H}}_{y^{*},d,h,q}}
\newcommand{\ghcn}{g^{\mathrm{H,CN}}_{y,q}}
\newcommand{\ghfp}{g^{\mathrm{H,FP}}_{y,q}}
\newcommand{\ghfpold}{g_{y-1,q}^{\mathrm{H,FP}}}
\newcommand{\ghcfd}{g^{\mathrm{H,CfD}}_{y,q}}
\newcommand{\ghcg}{g^{\mathrm{H,CG}}_{y,q}}
\newcommand{\ghinv}{g^{\mathrm{H,INV}}_{y,q}}
\newcommand{\deh}{d^{\mathrm{H}}_{y,d,h,q}}
\newcommand{\dng}{d^{NG}_{y,d,q}}
\newcommand{\dngstar}{d^{NG}_{y^{*},d,q}}
\newcommand{\ich}{IC^{\mathrm{H}}_{q}}
\newcommand{\cph}{cp^{\mathrm{H}}_{y,q}}
\newcommand{\cpht}{cp^{\mathrm{H,T}}_{q}}
\newcommand{\legcaph}{\overline{cp^{\mathrm{H}}_{y,q}}}
\newcommand{\cphstar}{cp^{\mathrm{H}}_{y^{*},q}}
\newcommand{\svh}{SV^\mathrm{H}_{y,p}}
\newcommand{\phc}{\lambda^{\mathrm{H,CN}}_{y}}
\newcommand{\bh}{b^{\mathrm{H}}_{y,q}}
\newcommand{\bhi}{b^{\mathrm{H,i}}_{y,q}}
\newcommand{\bhstar}{b^{\mathrm{H}}_{y^{*}}}
\newcommand{\png}{\lambda^{\mathrm{NG}}_{y}}
\newcommand{\sums}{\sum_{s \in \setS}}
\newcommand{\allq}{\forall q \in \setQ}
\newcommand{\alls}{\forall s \in \setS}
\newcommand{\lth}{LT^\mathrm{H}_{y,y^{*},q}}
\newcommand{\effelec}{\eta^{E\rightarrow H_2}_{q}}
\newcommand{\ghimp}{g^{\mathrm{H}}_{y,d,s}}
\newcommand{\ghimpcn}{g^{\mathrm{H,CN}}_{y,d,s}}
\newcommand{\cihimp}{CI^{\mathrm{H}}_{s}}
\newcommand{\pimp}{\lambda^{\mathrm{H,IMP}}_{y,d}}
\newcommand{\dhydrogen}{D^{\mathrm{H}}_{y,d}}
\newcommand{\hydrogentarget}{HT_{y}}
\newcommand{\hydrogentargetstar}{HT_{y^*}}
\newcommand{\hydrogentargetcap}{HT_{y}^{cap}}
\newcommand{\yoyf}{\Delta^{max.}_f}
\newcommand{\yoyq}{\Delta^{max.}_q}
\newcommand{\yoyi}{\Delta^{max.}_{I,y}}

% Command related to fossil based hydrogen production
\newcommand{\ghf}{g^{\mathrm{H}}_{y,d,f}}
\newcommand{\bhf}{b^{\mathrm{H}}_{y,f}}
\newcommand{\bhif}{b^{\mathrm{H,i}}_{y,f}}
\newcommand{\cphf}{cp^{\mathrm{H}}_{y,f}}
\newcommand{\effngf}{\eta^{NG\rightarrow H_2}_{f}}
\newcommand{\ichf}{IC^{\mathrm{H}}_{f}}
\newcommand{\lthf}{LT^\mathrm{H}_{y,y^{*},f}}
\newcommand{\legcaphf}{\overline{cp^{\mathrm{H}}_{y,f}}}
\newcommand{\cihf}{CI^{\mathrm{H}}_{f}}
\newcommand{\bhstarf}{b^{\mathrm{H}}_{y^{*},f}}
\newcommand{\svhf}{SV^\mathrm{H}_{y,f}}

% Command related to hydrogen import
\newcommand{\ghi}{g^{\mathrm{H}}_{y,d,I}}
\newcommand{\ghiold}{g^{\mathrm{H}}_{y-1,d,I}}

% Command related to long term contracts
\newcommand{\cdh}{CD^\mathrm{H}_{y,y^{*},q}}

% New commands
\newcommand{\phq}{\lambda^{\mathrm{H}}_{y}}
\newcommand{\ghq}{g^{\mathrm{H}}_{y,q}}
\newcommand{\ghcnq}{g^{\mathrm{H,CN}}_{y,q}}
\newcommand{\dngq}{d^{NG}_{y,q}}
\newcommand{\phcn}{\lambda^{\mathrm{capHCN}}_{y}}
\newcommand{\cphcn}{cp^{\mathrm{H,CN}}_{y,q}}
\newcommand{\dngstarq}{d^{NG}_{y^{*},q}}
\newcommand{\phrefq}{\lambda^{\mathrm{H},ref}_{y}}
\newcommand{\cphstarf}{cp^{\mathrm{H}}_{y^{*},f}}

\newcommand{\profh}{ P (\gh, \;  \cph)}

%%%%%%%%%%%%%%%%%%%%%%%%%%%%%%%%%%%%%%
%  Design of Carbon Contracts for Difference
%%%%%%%%%%%%%%%%%%%%%%%%%%%%%%%%%%%%%%

\newcommand{\piC}{\pi^C_y}
\newcommand{\piR}{\pi^R_y}
\newcommand{\piH}{\pi^H_y}
\newcommand{\piCNH}{\pi^{H,CN}_y}
\newcommand{\piHI}{\pi^{H,I}_y}

\newcommand{\setYstar}{\mathcal{Y}^*}
\newcommand{\allystar}{\forall y \in \setYstar}
\newcommand{\noallystar}{\forall y \notin \setYstar}
\newcommand{\sumystar}{ \sum_{\allystar} }
\newcommand{\gconv}{q^{conv}_y}
\newcommand{\gcn}{q^{cn}_y}
\newcommand{\gccfd}{q^{CCfD}}
\newcommand{\cconv}{ cap^{conv}_y}
\newcommand{\cpcn}{ cap^{cn}_y}
\newcommand{\Lap}{\mathcal{L}^{conv}}
\newcommand{\Lapcn}{\mathcal{L}^{cn}}
\newcommand{\bccfd}{B_{CCfD}}
\newcommand{\pstrike}{\lambda_{CCfD}}
\newcommand{\pref}{\lambda_{CO2,y}}
\newcommand{\prefh}{\hat{\lambda}_{CO2,y}}
\newcommand{\pgood}{\lambda_{g,y}}
\newcommand{\econv}{e_{conv}}
\newcommand{\eproj}{e_{proj}}
\newcommand{\bind}{b^{Ind}_y}
\newcommand{\bconv}{b^{conv}_y}
\newcommand{\emmis}{E_{ref} - (\frac{\pref}{\beta})^{1/\gamma}}
\newcommand{\vccn}{vc_y^{cn}}
\newcommand{\vconv}{vc^{conv}_y}
\newcommand{\vccnt}{vc_y^{cn}}
\newcommand{\vct}{vc^{conv}_y}
\newcommand{\vccnh}{\hat{vc}^{cn}_y}
\newcommand{\vconvh}{\hat{vc}^{conv}_y}
\newcommand{\ydagger}{y^{\dagger}}
\newcommand{\setYdagger}{\mathcal{Y}^{\dagger}}
\newcommand{\allydagger}{\forall y \in \setYdagger}
\newcommand{\noallydagger}{\forall y \notin \setYdagger}
\newcommand{\cpt}{\delta_{CPT}}

%%%%%%%%%%%%%%%%%%%%%%%%%%%%%%%%%%%%%%%%
%   Integrating heat in electrolyser operational models
%%%%%%%%%%%%%%%%%%%%%%%%%%%%%%%%%%%%%%%%

\newcommand{\pcomp}{p^{comp}_t}
\newcommand{\ptotal}{p^{total}_t}
\newcommand{\pcell}{p^{cell}_t}
\newcommand{\qstandby}{\dot{Q}^{standby}_t}
\newcommand{\qwater}{\dot{Q}^{water}_t}
\newcommand{\qcool}{\dot{Q}^{cool}_t}

%% file: appendix-A.tex
This appendix discusses the different electrochemical electrolyser models, describing the relation between the electrical power consumption, the operating temperature and the current density of the considered electrolyser types. Using Eq. \eqref{eq:P_cell}, the electrical power can be written as a function of temperature and current density.
\begin{equation}
    P_{cell} = U\cdot I
    \label{eq:P_cell}
\end{equation}
Electrolysis is no reversible process but has associated losses. The cell voltage can be represented by a reversible potential $U_{rev}$ and three overpotentials, as in Eq. \eqref{eq:Urev+overpotentials}: i) the activation overpotential $U_{act}$ represents the voltage required to overcome the activation energy of the chemical reaction; ii) the ohmic overpotential $U_{ohm}$ the voltage to overcome the finite electrical conductivity of the electrodes and electrolyte; and iii) the concentration overpotential $U_{conc}$ needed due to limitated mass transfer at the porous electrodes, which limit reaction kinetics \cite{buttler_22detailed_2015}.\footnote{This equation is valid for all three electrolyser types. However, the magnitudes of the overpotentials differ between technologies}
\begin{equation}
    U = U_{rev} + U_{act} + U_{ohm} + U_{conc}
    \label{eq:Urev+overpotentials}
\end{equation}
\subsection{Solid oxide electrolyser electrochemical model}
Using the Nernst equation, the reversible voltage can be written as a function of temperature and partial pressures, given in Eq. \eqref{eq:UrevSOEC} \cite{zhao_84system_2021}. Here $T$ is the temperature [K], $p_{H_2}, p_{O_2}$ and $p_{H_2O}$ are the partial pressures [bar] of $H_2, O_2$ and $H_2O$ respectively and $R$ is the universal gas constant [J/(mol K)]. The activation overpotential is given by \cite{buttler_22detailed_2015} \eqref{eq:Uact}. With the Faraday constant $F$ [C/mol], the current density $j$ [A/m$^2$]. The ohmic overpotential can be written as a function of the thickness $d_c$, $d_e$ and $d_a$  [m] and conductivity $\sigma_c$, $\sigma_e$ and $\sigma_a$ [$\Omega^{-1}$m$^{-1}$] of the cathode, electrolyte and anode respectively \cite{buttler_22detailed_2015, yin_64control-oriented_2022}.\footnote{The conductivity of the electrodes is assumed to be constant, while the electrolyte conductivity depends on the temperature and is given by $\sigma_{electrolyte}(T) = 33.4\cdot 10^3 exp(\frac{-10.3\cdot 10^3}{T})$ \cite{yin_64control-oriented_2022}.} The concentration overpotential depends on the rate at which reactants approach and products leave the electrodes, resulting in the dependency on the partial pressures of the reagents and reaction products shown in Eq. \eqref{eq:Uconc} \cite{zhao_84system_2021}. $D^{eff}_{H_2O}$ [m$^2$/s] represents the effective diffusion coefficient, $\mu$ [10$^{-7}$Pa$\cdot$s] the dynamic viscosity of oxygen and $B_g$ [m$^2$] the permeability.\footnote{Details on how these coefficients can be determined are given in \cite{zhao_84system_2021}}
\begin{align}
    U_{rev}^{SOE} = 1.253 - 2.4516 \cdot 10^{-4}\cdot T \nonumber \\
    + \frac{RT}{2F}ln\bigg(\frac{p_{H_2}\cdot (p_{O_2})^{1/2}}{p_{H_2O}}\bigg)
    \label{eq:UrevSOEC}
\end{align}
\begin{align}
    U_{act}^{SOE} = \frac{RT}{F}sinh^{-1}\bigg(\frac{j}{2\gamma_i \cdot exp(-\frac{E_{act,a}}{RT})}\bigg) \nonumber \\
    + \frac{RT}{F}sinh^{-1}\bigg(\frac{j}{2\gamma_i \cdot exp(-\frac{E_{act,c}}{RT})}\bigg)
    \label{eq:Uact}
\end{align}
\begin{equation}
    U_{ohm}^{SOE} = j\bigg(\frac{d_{c}}{\sigma_{c}}+\frac{d_{e}}{\sigma_{e}(T)}+\frac{d_{a}}{\sigma_{a}}\bigg)
    \label{eq:Uohm}
\end{equation}
\begin{align}
     U_{conc}^{SOE} = \frac{RT}{2F}ln\bigg(\frac{1+\frac{jRTd_{c}}{2Fp_{H_2}D^{eff}_{H_20}}}{1-\frac{jRTd_{c}}{2Fp_{H_2O}D^{eff}_{H_20}}}\bigg) \nonumber \\
     +\frac{RT}{4F}ln\bigg(\frac{\sqrt{(p_{O_2}^2+\frac{jRT\mu d_{a}}{2FB_g}}}{p_{O_2}}\bigg)
     \label{eq:Uconc}
\end{align}  
\subsection{PEM electrolyser electrochemical model}
The Nernst equation can be used to represent the reversible overpotential of the PEM electrolyser as well, albeit again in a slightly adapted form shown in Eq. \eqref{eq:UrevPEM} \cite{abdin_42modelling_2015}. For the PEM electrolyser, there is no aqueous $KOH$ solution since the electrolyte is a solid. Because of the pure water, the water activity $a_w$ is equal to one \cite{abdin_42modelling_2015}. The activation overpotential can be represented by Eq. \eqref{eq:UactPEM} \cite{chitsaz_113thermodynamic_2019}. As was the case for the AWE, $\alpha$ represents the electrons' transfer coefficient. For the PEM electrolyser, it is reasonable to assume that $\alpha$ is equal to 0.5 for both the cathode and anode \cite{abdin_42modelling_2015}. \cite{carmo_105comprehensive_2013} discuss the wide range of values for the exchange current densities of a PEM electrolyser appearing in the literature. For the anode exchange current density, values between $10^{-12}$ A/m$^2$ and $1.3 \cdot 10^{-3}$ A/m$^2$ are reported, while the values for the cathode exchange current density can range from $10^{-3}$ A/m$^2$ to $1.8 \cdot 10^{-1} $ A/m$^2$. On average, the cathode exchange current density is a factor $10^4$ larger than the anode exchange current density. The influence of the catalyst, temperature, electrode material and porosity again constitute the wide range of values present in the literature. To be able to include the effect of temperature on the exchange current densities, the following equations are used, with $j_{0,ref}$ equal to $1.08\cdot10^{-17}$ A/m$^2$ \cite{chitsaz_113thermodynamic_2019}. The ohmic overpotential depends on the thickness $d$ and conductivity  $\sigma$ of the different elements  \cite{abdin_42modelling_2015}.\footnote{The conductivity of the electrode $\sigma_{electrode}$ in $\Omega^{-1}$m$^{-1}$ is assumed to be constant, whereas the membrane conductivity $\sigma_{membrane}$ depends on the temperature: $\sigma_{membrane}(T)= (0.0439T-3.8084) \cdot exp(1268(\frac{1}{303}-\frac{1}{T}))$ \cite{abdin_42modelling_2015, hernandez-gomez_66investigation_2020}.} The concentration overpotential is given by Equation \eqref{eq:UconcAWE} \cite{firtina-ertis_95thermodynamic_2022}, where $j_{L}$ is the limiting current density.\footnote{To prevent the overpotential from going to infinity at the maximum current density, the limiting current density is often set to $1.05\cdot j_{max}$ \cite{firtina-ertis_95thermodynamic_2022}.}
\begin{align}
    U_{rev}^{PEM} = 1.229 - 0.9\cdot 10^{-3} \cdot (T-298) \nonumber \\
    + \frac{RT}{2F} \cdot ln\bigg(\frac{p_{H_2}\cdot p_{O_2}^{1/2}}{a_w}\bigg)
    \label{eq:UrevPEM}
\end{align}
\begin{align}
    U_{act}^{PEM} = \frac{RT}{\alpha 2 F}ln\bigg(\frac{j}{j_{0,ref} \cdot 10^4 \cdot exp(0.086T)}\bigg) \nonumber \\
    + \frac{RT}{\alpha 2F}ln\bigg( \frac{j}{j_{0,ref} \cdot exp(0.086T)}\bigg)
    \label{eq:UactPEM}
\end{align}
\begin{equation}
    U_{ohm}^{PEM} = \bigg(\frac{d_{electrode}}{\sigma_{electrode}} + \frac{d_{membrane}}{\sigma_{membrane}(T)}\bigg)
\end{equation}
\begin{equation}
    U_{conc}^{PEM} = -\frac{2R T}{F}\cdot ln \bigg(1-\frac{j}{j_{L}}\bigg)
    \label{eq:UconcAWE}
\end{equation}

%% file: appendix-B.tex
\subsection{Parameters}
\label{app:parameters}
\begin{table}[tbph]
    \centering
    \small
    \begin{tabular}{llll}
        \toprule
        Symbol & Value   & Symbol & Value    \\
        \midrule
         $E_{act,c}$ & $10^5$ J/mol & $E_{act,a}$ & $1.2\cdot 10^5$ J/mol  \\
        $p_{H_2}$ & $0.5$ bar  & $d_{cath}$ & $50\cdot 10^{-5}$ m  \\
        $p_{O_2}$ & $1$ bar  & $d_{anode}$ & $50\cdot 10^{-6}$ m  \\
        $p_{H_2O}$ & $0.5$ bar & $d_{elec}$ & $50 \cdot 10^{-6}$ m  \\
        $\gamma_c$ & $1.3 \cdot 10^{10}$ A/m$^2$  & $\sigma_{cath}$ & $8.4\cdot 10^{3} / (\Omega . \si{m})$ \\
        $\gamma_a$ & $2.1$ A/m$^2$  & $\sigma_{anode}$ & $8\cdot 10^4  / (\Omega . \si{m})$  \\
        \bottomrule
    \end{tabular}
        \caption{Parameters used in the SOE electrochemical model \cite{buttler_22detailed_2015, zhao_84system_2021,yin_64control-oriented_2022 }}
    \label{tab:combined_parameters}
\end{table}
\begin{table}[tbph]
    \centering
    \small
    \label{tab:pem_electrolyser_parameters}
    \begin{tabular}{llllll}
        \toprule
        Symbol & Value & Symbol & Value &  \\
        \midrule
          $\alpha$ & 0.5  & $j_{0,\text{ref}}$ & $1.08 \cdot 10^{-17}$ A/m$^2$  \\
        $p_{H_2}$ & $1$ bar  & $d_{\text{membrane}}$ & $2.54\cdot 10^{-4}$ m  \\
        $p_{O_2}$ & $1$ bar & $d_{\text{electrode}}$ & $8\cdot 10^{-4}$ m  \\
        $a_w$ & $1$ bar & $\sigma_{\text{electrode}}$ & $5.53\cdot 10^{6}$ ($\Omega\cdot$ m)$^{-1}$ \\
        \bottomrule
    \end{tabular}
        \caption{Parameters used in the PEM electrolyser electrochemical model \cite{zhao_84system_2021, chitsaz_113thermodynamic_2019, abdin_42modelling_2015}}
\end{table}
\begin{table}[tbph]
    \centering
    \small
    \begin{tabular}{llll}
        \toprule
        Symbol & Value & Symbol & Value \\
        \midrule
        $\Delta t$ & 1h & $M$ & $10^7$ \\
        $\eta_{EH}$ & 0.95 \cite{petipas_72benefits_2014} & $c_p^{water}$ & 4184 J/(kg$\cdot$K) \\
        $\eta_{ST}$ & 0.45 \cite{buttler_22detailed_2015} & $c_p^{steam}$ & 2323 J/(kg$\cdot$K) \\
        $M_{H_2}$ & 2.016$\cdot 10^{-3}$ kg/mol & $\Delta H$ & $2.256\cdot 10^6$ J/kg \\
        $M_{H_2O}$ & $18.016\cdot 10^{-3}$ kg/mol & $T_a$ & 293 K \\
        $\overline{p}$ & $20\cdot 10^6$ W & $\alpha$ & 2.92$\cdot 10^6$ W$\cdot$s/kg \\
        $A$ & 0.21 m$^2$ \\
        \bottomrule
    \end{tabular}
    \caption{General parameters used in all operational models}
    \label{tab:generalparametersoperational}
\end{table}
\begin{table}[tbph]
    \centering
    \small
    \begin{tabular}{lllll}
        \toprule
        Symbol & Value  & Symbol & Value  \\
        \midrule
        $n_c$ & 5776  & $\overline{T}$ & 1273 K  \\
        $C_h$ & $173.28\cdot 10^6$ J/K  & $\underline{T}$ & 1073 K \\
        $R_t$ & $1.3067\cdot 10^{-3}$ K/W  & $\overline{j}$ & 10000 A/m$^2$ \\
        $U_{tn}$ & 1.2995 V  & $\underline{j}$ & 2000 A/m$^2$ \\
        \bottomrule
    \end{tabular}
    \caption{Parameters used in SOE operational model}
    \label{tab:merged_parameters}
\end{table}
\begin{table}[tbph]
    \centering
    \small
    \begin{tabular}{llll}
        \toprule
        Symbol & Value  & Symbol & Value  \\
        \midrule
        $n_c$ & 1532 & $\overline{T}$ & 373 K  \\
        $\dot{m}^H_{\underline{T},\underline{I}}$ & $29.1\cdot 10^{-3}$ kg/s & $\underline{T}$ & 293 K \\
        $p^E_{\underline{T},\underline{I}}$ & $5.9\cdot 10^6$ W & $\overline{j}$ & 20000 A/m$^2$  \\
        $C_h$ & $45.96\cdot 10^6$ J/K  & $\underline{j}$ & 1500 A/m$^2$  \\
        $R_t$ & $1.067\cdot 10^{-4}$ K/W  & $U_{tn}$ & 1.4813 V  \\
        \bottomrule
    \end{tabular}
        \caption{Parameters used in PEM operational model}
    \label{tab:pem_parameters}
\end{table}
\begin{table}[tbph]
\centering
\small
\begin{tabular}{>{\centering\arraybackslash}llll}
\hline
Symbol  & SOE  & PEM & Unit \\ \midrule
$a_{1,1}$ &  -0.926   &  -8.253  &  W/K\\
$a_{2,1}$ &  -2.873    &   -28.833  &  W/K\\
$a_{1,2}$ &  -0.385   &  -6.942  &       W/K\\
$a_{2,2}$ &   -1.199  & -20.188   &       W/K\\
$b_{1,1}$ & 0.285    &  0.517   &      W/(A/m$^2$)\\
$b_{2,1}$ & 0.327     & 0.675   &      W/(A/m$^2$)\\
$b_{1,2}$ &  0.262   &  0.464   &    W/(A/m$^2$)\\
$b_{2,2}$ &  0.284    &  0.588  &      W/(A/m$^2$)\\
$c_{1,1}$ & 968.642     & 2453.652      & W\\ 
$c_{2,1}$ & 2906.471     & 7185.014      & W\\ 
$c_{1,2}$ & 431.063    & 2352.914       & W\\ 
$c_{2,2}$ & 1290.996     & 5684.778       & W\\ 
\midrule
\end{tabular}
\caption{Coefficients of the piecewise linearisation of the power curve of each electrolyser type using four segments}
\end{table}

%% file: appendix-C.tex
\subsection{Extra results case study}
\label{app:extra results}
\begin{figure}[tbph]
\begin{subfigure}{0.45\textwidth}
  \centering
  \includegraphics[width=\linewidth]{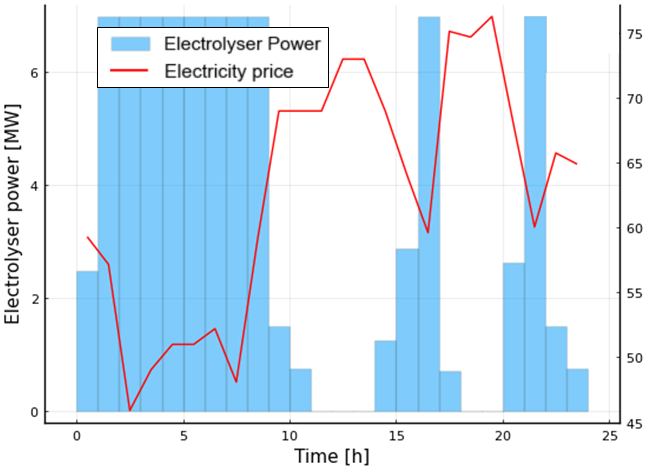}
  \caption{}
  \label{fig:PEMpowerprice}
\end{subfigure}
\begin{subfigure}{.45\textwidth}
  \includegraphics[width=\linewidth]{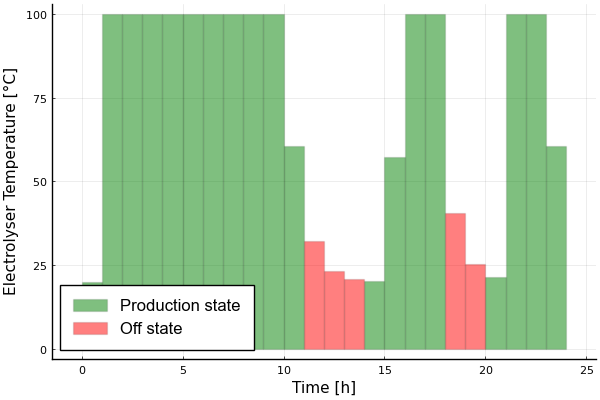}
  \caption{}
  \label{fig:PEMtempbarscolours}
\end{subfigure}%
\caption{(a) PEM power and electricity price and (b) temperature and operational states, for 5th Jan 2019 using 3.5 EUR/kg $\si{H_2}$}
\end{figure}
\begin{table}[H]
\centering
\begin{tabular}{cc|c|c|c|}
\cline{3-5}
                                                                     &    & \makecell{Average \\ profit/day \\ {[EUR]}} & \makecell{Hours/y in \\production\\ state [h]} & \makecell{Required \\ profit/day \\ {[EUR]}}  \\ \hline
\multicolumn{1}{|c|}{\multirow{4}{*}{\makecell{2.5 EUR/kg $\si{H_2}$}}} & Model 1) & 9053                        & 8473                          &  41479                                  \\ 
\multicolumn{1}{|l|}{}                                               & Model 2) & 10597                       & 8616                          & 42155                                    \\ 
\multicolumn{1}{|l|}{}                                               & Model 3) & 11158                       & 8600                          & 42079                                       \\ \hline
\multicolumn{1}{|c|}{\multirow{4}{*}{\makecell{3.5 EUR/kg $\si{H_2}$}}} & Model 1) & 19763                       & 8742                          & 42750                                   \\ 
\multicolumn{1}{|l|}{}                                               & Model 2) & 21431                       & 8754                          & 42807                                      \\  
\multicolumn{1}{|l|}{}                                               & Model 3) & 21999                       & 8753                          & 42802                                    \\ \hline
\multicolumn{1}{|c|}{\multirow{4}{*}{\makecell{4.5 EUR/kg $\si{H_2}$}}} & Model 1) & 30694                       & 8760                          & \multicolumn{1}{c|}{}                   \\ 
\multicolumn{1}{|l|}{}                                               & Model 2) & 32374                       & 8760                          & \multicolumn{1}{c|}{}                   \\ 
\multicolumn{1}{|l|}{}                                               & Model 3) & 32943                       & 8760                          & \multicolumn{1}{c|}{}                   \\ \hline
\multicolumn{1}{|c|}{\multirow{4}{*}{\makecell{5.5 EUR/kg $\si{H_2}$}}} & Model 1) & 41642                       & 8760                          & \multicolumn{1}{c|}{}                   \\ 
\multicolumn{1}{|l|}{}                                               & Model 2) & 4332                        & 8760                          & \multicolumn{1}{c|}{}                   \\ 
\multicolumn{1}{|l|}{}                                               & Model 3) & 43891                       & 8760                          & \multicolumn{1}{c|}{}                   \\ \hline
\end{tabular}
\caption{Results for a 15 MW SOE for different hydrogen prices, where the following notation is used: Model 1) SOE without heat integration, Model 2) SOE with low-temperature heat integration, Model 3) SOE with high-temperature heat integration}
\label{tab:resultsSOE}
\end{table}

\begin{table}[tbph]
\centering
\begin{tabular}{cc|c|c|c|}
\cline{3-5}
                                                                     &    & \makecell{Average \\ profit/day \\ {[EUR]}} & \makecell{Hours/y in \\production \\state [h]} & \makecell{Required \\ profit/day \\ {[EUR]}}\\ \hline
\multicolumn{1}{|c|}{\multirow{2}{*}{\makecell{ 2.5 EUR/kg $\si{H_2}$}}}                
                                                                     & Model 4) & 2040                       & 7255        &3677-5544                  \\ %\cline{2-5} 
\multicolumn{1}{|l|}{}                                               & Model 5) & 2077                     & 7114       &3616-5447                 
                        \\ \hline
\multicolumn{1}{|c|}{\multirow{2}{*}{\makecell{3.5 EUR/kg $\si{H_2}$}}} 
                                                                     & Model 4) & 6227                       & 8578     &4243-6455                   \\ %\cline{2-5} 
\multicolumn{1}{|l|}{}                                               & Model 5) & 6288                    & 8547    &4230-6434                    
                          \\ \hline
\multicolumn{1}{|c|}{\multirow{2}{*}{\makecell{ 4.5 EUR/kg $\si{H_2}$}}}
                                                                     & Model 4) & 11594                     & 8740     &4313-6567                    \\ %\cline{2-5}
\multicolumn{1}{|l|}{}                                               & Model 5) & 11667                       & 8737    &4311-6564                     
                      \\ \hline
\multicolumn{1}{|c|}{\multirow{2}{*}{\makecell{ 5.5 EUR/kg $\si{H_2}$}}}
                                                                     & Model 4) & 17298                      & 8759             &4321-6580                        \\ %\cline{2-5}
\multicolumn{1}{|l|}{}                                               & Model 5) & 17373                      & 8758   &4230-6579                      
                          \\ \hline
\end{tabular}
\caption{Results for a 15 MW PEM electrolyser for different hydrogen prices, where the following notation is used: Model 4) PEM without heat integration, Model 5) PEM with heat integration}
\label{tab:resultsPEM}
\end{table}